\documentclass[useAMS,usenatbib]{mn2e}

\usepackage[dvipdfmx]{graphicx}
\usepackage{color}
\usepackage{amsmath}
\usepackage{journalnames}

\usepackage[T1]{fontenc}
\usepackage{aecompl}
\usepackage{yfonts}

\newcommand{\lesssim}{\,\raisebox{-0.4ex}{$\stackrel{<}{\scriptstyle\sim}$}\,}

\begin{document}


\title[Simulating IFU observations of globular clusters]{Understanding the central kinematics of globular clusters with simulated integrated-light IFU observations}

\author[P. Bianchini et al.]{Paolo Bianchini\thanks{E-mail:
bianchini@mpia.de}\thanks{Member of the International Max Planck Research School for Astronomy and Cosmic Physics at the University of Heidelberg, IMPRS-HD, Germany.}, Mark A. Norris, Glenn van de Ven, Eva Schinnerer\\
Max-Planck Institute for Astronomy, Koenigstuhl 17, 69117 Heidelberg, Germany\\
}

\date{Accepted for publication in MNRAS}
\pagerange{} \pubyear{}
\maketitle
\label{firstpage}


\begin{abstract}
The detection of intermediate mass black holes in the centres of globular clusters is highly controversial, as complementary observational methods often deliver significantly different results.
In order to understand these discrepancies, we develop a procedure to simulate integral field unit (IFU) observations of globular clusters: Simulating IFU Star Cluster Observations (SISCO). The input of our software are realistic dynamical models of globular clusters that are then converted in a spectral data cube.
We apply SISCO to Monte Carlo cluster simulations from \citet{Downing2010}, with a realistic number of stars and concentrations. Using independent realisations of a given simulation we are able to quantify the stochasticity intrinsic to the problem of observing a partially resolved stellar population with integrated-light spectroscopy. 
We show that the luminosity-weighted IFU observations can be strongly biased by the presence of a few bright stars that introduce a scatter in the velocity dispersion measurements up to $\simeq40\%$ around the expected value, preventing any sound assessment of the central kinematic and a sensible interpretation of the presence/absence of an intermediate mass black hole.
Moreover, we illustrate that, in our mock IFU observations, the average kinematic tracer has a mass of $\simeq0.75$ $M_\odot$, only slightly lower than the mass of the typical stars examined in studies of resolved line-of-sight velocities of giant stars. Finally, in order to recover unbiased kinematic measurements we test different masking techniques that allow us to remove the spaxels dominated by bright stars, bringing the scatter down to a level of only a few percent. The application of SISCO will allow to investigate state-of-the-art simulations as realistic observations.

\end{abstract}


\section{Introduction}

The study of the internal kinematics of globular clusters (GCs) offers the possibility of unveiling the complexity of these stellar systems, previously regarded as simple, spherical and isotropic. In fact, kinematic measurements are key to understand the formation and dynamical evolution of GCs, providing insights into the role of ingredients such as internal rotation, velocity anisotropy, presence of intermediate mass black holes (IMBHs) in the centre of GCs, fundamental in shaping their internal structure. 

In particular, significant effort has been devoted to the search for IMBHs, postulated to have a mass intermediate between those of stellar mass black holes (M$_\bullet<100$ M$_\odot$) and those of the super-massive black holes (SMBH, M$_\bullet>$ 10$^5$ M$_\odot$) found at the centres of galaxies.
These putative intermediate mass black holes (IMBHs) have proven elusive, with disputed evidence for their presence in Galactic globular clusters (GCs, \citealp{Gebhardt2000,vandenBosch2006,Noyola2010,vanderMarel2010,Luetzgendorf2013,Lanzoni2013,denBrok2014}) and extra-galactic ultra-luminous X-ray sources (e.g., \citealp{Matsumoto2001,Fabbiano2001}).
The search for the existence of IMBHs within GCs has partially been motivated by the observation that the extrapolation of the $M_\bullet-\sigma$ relation for galaxies, linking the 
mass of the central back hole to the velocity dispersion of the host stellar system \citep{Ferrarese2000,Magorrian1998}, suggests that GCs are ideal environments 
to find central IMBHs with masses of $10^3-10^4$~M$_\odot$. 

The search for IMBHs reported in the literature is based primarily on two channels: detection of radio and X-ray emission \citep{Miller2002, Maccarone2008,Strader2012,Kirsten2012}, and detection of kinematic signatures in the central region of GCs, such as the rise of the central velocity dispersion (e.g. \citealp{Bahcall1976,Luetzgendorf2013}). The detectability of the black hole by the former method is dependent on the feeding of the black hole with gas, an event that is inefficient in the extremely gas poor environment of most GCs.The latter method requires instead very precise velocity measurements (accuracy of the order of 1 km s$^{-1}$ to reliably measure central velocity dispersions of $\simeq10$ km s$^{-1}$) with high spatial resolution of the very crowded central region of GCs (central few arcseconds).

Two main strategies are generally used to acquire the necessary kinematic information of the central region of GCs: 1) resolved kinematics, by measuring discrete velocities 
of individual stars (using line-of-sight velocities or proper motions), 2) unresolved kinematics, by measuring the velocity dispersion from line broadening of integrated light spectra with either integral field units (IFUs) or slit spectroscopy (e.g, \citealp{Dubath1997}). Unsettlingly, these apparently complementary methods can give significantly different observational outcomes when applied to the same object, making the detection of IMBHs highly ambiguous.
 
In particular, integrated light spectroscopy seems to measure rising central velocity dispersions, favouring the presence of IMBHs (see e.g., \citealp{Noyola2010} for 
$\omega$ Cen; \citealp{Luetzgendorf2011,Luetzgendorf2015} for NGC 6388), while resolved stellar kinematics do not confirm the presence of this signature (see \citealp{vanderMarel2010} for 
proper motion measurements of $\omega$ Cen; \citealp{Lanzoni2013} for individual line-of-sight measurements in NGC 6388).

Therefore, understanding the possible sources of biases affecting the different methods is an essential first step to undertake before interpreting any 
kinematic signatures possibly connected to the presence of IMBHs. The difference between the techniques can arise because unresolved measurements give 
intrinsically luminosity-weighted kinematic information, whereas in resolved kinematic studies the kinematic profiles are constructed assigning to each discrete measurement the same weight. Progress in understanding how this difference could influence the measurements has been made, for example, in understanding that unresolved kinematics 
can be strongly biased by the presence of a few bright stars dominating the integrated spectra, increasing the shot noise of the velocity dispersion (e.g. see \citealp{Dubath1997,Noyola2010,Lanzoni2013}). Moreover, \citet{Luetzgendorf2015} recently showed that measurements of velocity dispersions from discrete velocities of individual stars can also be biased towards lower values of velocity dispersions.

In order to undertake an exploration of the issues that emerge from applying integrated-light spectroscopy to systems with a (partially) resolved stellar population, like Galactic GCs, we develop a new procedure to simulate IFU observations of globular clusters: Simulating IFU Star Cluster Observations (SISCO). Starting from realistic Monte Carlo cluster simulations from \citet{Downing2010} (for details on the Monte Carlo code see \citealp{Giersz1998,Hypki2013,Giersz2013}), the output of our simulation is a data cube with spectra for every pixel in a selected field-of-view. Our work is motivated by the growing use of such techniques to study the general kinematic properties of Galactic GCs (see for example the study of internal rotation by \citealp{Fabricius2014} in addition to the applications related to IMBHs mentioned above). We focus in particular on investigating the bias due to stochastic effects and shot noise introduced by a few bright stars, and explore possible physical interpretations (e.g. mass segregation) of the discrepancies reported in the literature (e.g. those relative to NGC 6388, \citealp{Luetzgendorf2011,Lanzoni2013}).

The paper is organised as follows. In Section 2, we introduce the properties of the Monte Carlo cluster simulations that we will use in the rest of our work. In Section 3, we describe the method we have developed to construct IFU mock observations starting from a general cluster simulation. In Section 4, we analyse the kinematics of our mock observations and investigate the possible biases intrinsic to luminosity-weighted kinematic measurements. In Section 5, we outline and thoroughly test masking techniques to minimise stochastic scatter and to recover unbiased velocity dispersion measurements. Finally, in Section 6, we present our conclusions and future prospectives.

\section{Simulations of globular clusters}
\label{sec:MCsim}
\begin{table*}
\begin{center}
\caption{Structural properties of the Monte Carlo cluster simulations used in this work, for a snapshot at 13 Gyr.}
\begin{tabular}{llrrr}
\hline\hline
Simulation properties& &Simulation A& Simulation B& Simulation C\\
\hline
Number of particles &$N$&$1.8\times10^6$&  $4.1\times10^5$& $1.8\times10^6$\\
Total mass& $M_\mathrm{tot}$& $6.7\times10^5$ $M_\odot$&$1.5\times10^5$ $M_\odot$&$6.7\times10^5$ $M_\odot$\\
Distance & d & 10 kpc &10 kpc&20 kpc\\
Half-light radius& $R_h$&2.8 pc / 57.8 arcsec&2.24 pc / 46.2 arcsec &2.8 pc / 28.9 arcsec\\
Core radius & $R_c$&1.3 pc / 26.8 arcsec&0.39 pc / 8.0 arcsec&1.3 pc / 13.4 arcsec\\
Concentration& $C=\log (R_t/R_c)$& 1.6& 2.3 &1.6\\
Central luminosity density & $l_0$ & 64.4 L$_\odot$ arcsec$^{-2}$&37.5 L$_\odot$ arcsec$^{-2}$&234.7 L$_\odot$ arcsec$^{-2}$\\
Binary fraction&$f_b$&4\%&4\%&4\%\\
Metallicity &$[Fe/H]$& -1.3&-1.3& -1.3\\
\hline
\end{tabular}
\label{tab:MCsimulations}
\end{center}
\end{table*}

The starting point of our work are Monte Carlo cluster simulations, developed by \citet{Downing2010} with the Mocca Monte Carlo code \citep{Giersz1998,Hypki2013}. These simulations were not originally designed for our study. However, they provide the realistic long term dynamical evolution of globular clusters with a single stellar population, in which ingredients such as stellar mass function, stellar evolution, initial binary fraction are taken into consideration, and therefore are suitable to test the performance of our tool SISCO (see Sect. \ref{SISCO}).

The main set of simulations (labeled as Simulation A in Tab. \ref{tab:MCsimulations}) starts with $2\times10^6$ particles drawn from a \citet{Plummer1911} model, with a \citet{Kroupa2001} initial mass function, 10\% primordial binary fraction, metallicity [Fe/H]=-1.3, and they are evolved in isolation.\footnote{The assumption of isolation is not a limitation since we are only interested in the central region of clusters, where any effect of an external tidal field on a GC would be negligible.} The simulations have no central intermediate mass black hole and internal rotation is not considered (note however that internal rotation is observed in several GCs, e.g. \citealp{Bianchini2013, Fabricius2014, Kacharov2014,Lardo2015}). Five different independent realisations of the same simulation are available, and we will use them to analyse stochastic effects. At 13 Gyr, the simulations are characterised by $N\simeq1\,800\,000$ particles (both single and binary stars), $\simeq4\%$ binary fraction, total mass of $M_\mathrm{tot}\simeq6.7\times10^5$ $M_\odot$, projected half light radius of $R_h\simeq2.8$ pc, core radius of $R_c\simeq$1.3 pc and concentration $C=\log(R_t/R_c)\simeq1.6$, with $R_t$ tidal radius of the cluster. Moreover, the resulting clusters are isotropic in the central regions and mildly radially anisotropic in the outer parts. Having included a stellar mass function, they are also characterized by dynamical mass segregation as described in Sect. \ref{sec:tracers} (no initial mass segregation is assumed). For a summary of the structural properties, see Tab. \ref{tab:MCsimulations}.

We place the simulated globular clusters at 10 kpc from the observer with a global systemic line-of-sight velocity of 300 km s$^{-1}$, to match the typical properties of a Galactic GC. The data from the Monte Carlo simulations that we will need are: the spatial coordinates of the stars in the plane of the sky, the velocity of each star along the line-of-sight, the stellar parameters of each star (effective temperature $T_\mathrm{eff}$, mass $m_\star$, luminosity $l_\star$) and the metallicity of the cluster. Any other globular cluster simulation providing this information can also be used in the software package that we have developed.

We further consider two additional sets of simulations: (i) a more concentrated simulation (labeled Simulation B), with $N\simeq410\,000$ stars, concentration c=2.3, core radius $R_c=8$ arcsec, observed at 10 kpc, (ii) and a more crowded simulation (labeled Simulation C), obtained placing Simulation A at a distance of 20 kpc. The former simulation, even if more centrally concentrated, does not represent a more crowded case, since Simulation B is performed with fewer stars than Simulation A. The latter simulation has a higher number of giants stars in the FoV, making the crowding $\approx4$ times higher than Simulation A (see the central luminosity density in Tab. \ref{tab:MCsimulations}). In the following section we will consider Simulation A, and we will discuss the results connected to Simulation B and C in Sect. \ref{sec:masking}.
\section{Simulations of IFU observations}
\label{SISCO}
In this section we describe step-by-step our procedure, Simulating IFU Star Cluster Observations (SISCO), to build a mock IFU observation of a globular cluster, starting from the 
Monte Carlo cluster simulations described in the previous section. The final product will consist of IFU simulations in the Calcium triplet wavelength range (8400-8800 \AA) with spectra associated to every spaxel. 
From this, we will build kinematic maps and kinematic profiles in the same manner as observers.

\subsection{From stellar parameters to stellar spectra}
\begin{figure}
\centering
\includegraphics[width=0.5\textwidth]{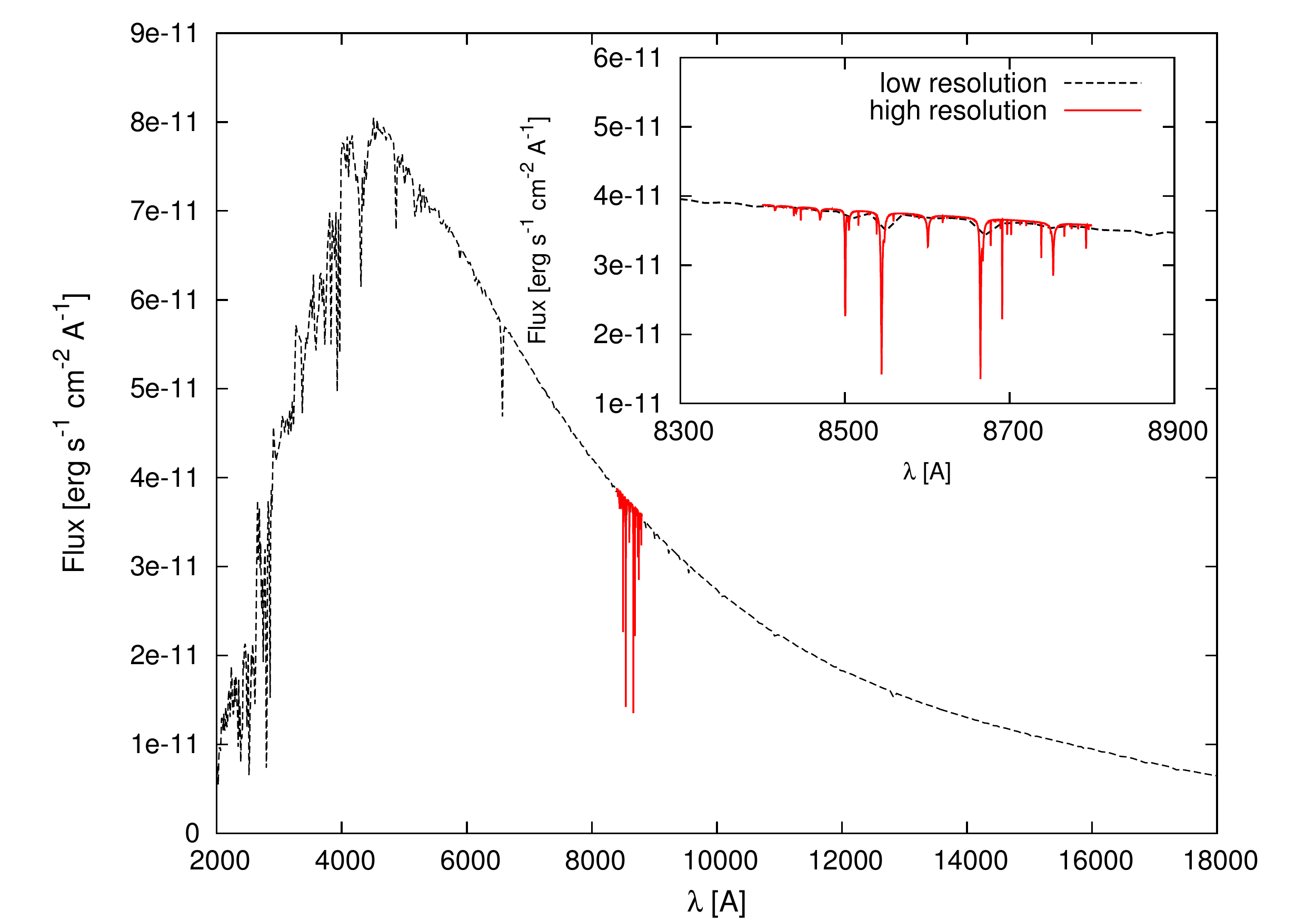}
\caption{Stellar spectrum associated to a main sequence star with $T_\mathrm{eff}=5815$, $m_\star=0.72$ M$_\odot$ and $l_\star$=$0.66$ L$_\odot$ using the GALEV evolutionary synthesis model. The corresponding high resolution synthetic spectrum (from the MARCS synthetic stellar library) in the region of the CaII-triplet is shown with the red line. With a resolving power of $R=20\,000$, these spectra are ideal for measurements of the internal kinematics of globular clusters.}
\label{fig:spectrum}
\end{figure}

\begin{figure*}
\centering
\includegraphics[width=0.34\textwidth]{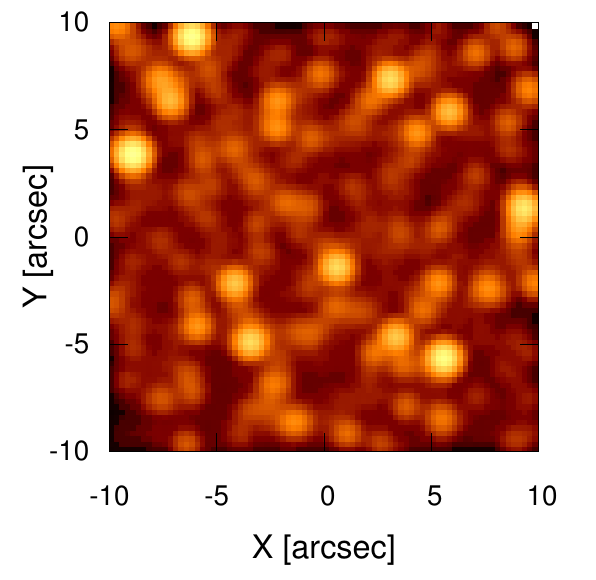}\,
\includegraphics[width=0.60\textwidth]{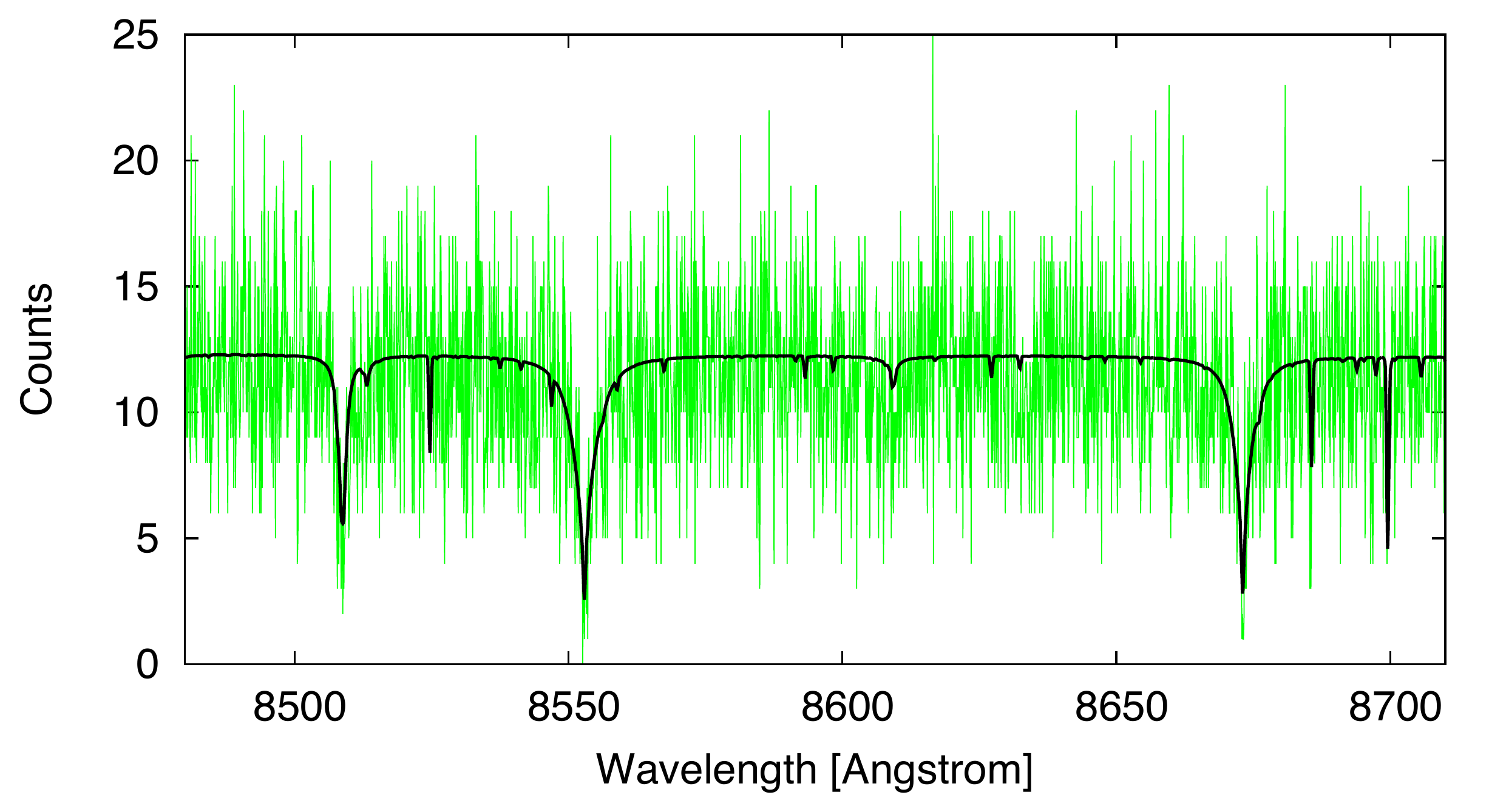}
\caption{\textbf{Left panel:} Luminosity map in logarithmic scale of the central $20\times20$ arcsec$^2$ region of our simulated globular cluster (Simulation A) placed at 10 kpc and observed with a seeing of 1 arcsec and an average signal-to-noise ratio of $S/N\simeq10$ per {\AA}. The luminosity map is constructed summing up all the flux from the spectra within each spaxel, in the 8400-8800 {\AA} range. \textbf{Right panel:} typical spectrum of a spaxel, obtained summing all the Doppler-shifted spectra falling in the spaxel, properly weighted by their PSF. The black line indicates the spectrum without noise, while the green line indicates the case of an observation with $S/N\simeq10$~per~{\AA}.}
\label{fig:map_lum}
\end{figure*}

The first step in building a IFU mock observation consists in associating a stellar spectrum to each star in a globular cluster simulation. We do this in two steps: first we associate to 
each star a low-resolution spectrum, covering a broad-wavelength, using the GALEV evolutionary synthesis model \citep{Kotulla2009}. GALEV is based on the BaSeL library of model 
atmospheres \citep{Lejeune1997, Lejeune1998a,Lejeune1998b} and gives stellar spectra from the extreme ultraviolet to the far infrared (9$-$160\,000 nm) with variable resolution ranging 
from 2 {\AA} to 400 {\AA} (20 {\AA} in the optical). In order to associate to each simulated star a stellar spectrum, the effective temperature $T_\mathrm{eff}$, mass $m_{\star}$, luminosity 
$l_{\star}$, and metallicity $Z$ outputs from the Monte Carlo simulations are matched to the appropriate GALEV spectrum. The computed spectra, in units of erg~cm$^{-2}$~s$^{-1}$~{\AA}$^{-1}$, 
can then be convolved with filter transmission curves to obtain the colour and magnitude information for every star (to build, for example, a colour-magnitude diagram in the desired filters; see Fig. \ref{fig:segregation}).

The second step consists of associating an additional high-resolution spectrum in the wavelength range that will be used in our mock observations. These further spectra are necessary 
in order to have high enough spectral resolution to measure the internal kinematics of globular clusters with typical velocity dispersions of 10 km s$^{-1}$. We select the wavelength range around the Calcium triplet (8400-8800 {\AA}) typically used to measure the kinematics in GCs, for example through spectroscopy with VLT/FLAMES.\footnote{Note that in principle, any other wavelength range can be used.} The high-resolution spectra are taken from the MARCS synthetic stellar library \citep{Gustafsson2008} and provide a constant resolving power $R=\lambda/\Delta\lambda=20\,000$, corresponding to a velocity dispersion of 15 km s$^{-1}$ around the Calcium triplet.

Once a high-resolution spectrum is associated to each star according to the stellar parameters, it is rescaled to match the flux of the low-resolution spectrum. Fig. \ref{fig:spectrum} shows both the low- and high-resolution spectra for a main sequence star with $T_\mathrm{eff}=5\,815$, $m_\star=0.72$ M$_\odot$ and $l_\star$=$0.66$ L$_\odot$. The correctly rescaled high resolution spectra will be used in the rest of our simulations. We finally proceed to Doppler-shift the single star spectra using the line-of-sight velocity given by our Monte Carlo cluster simulation. In the case of binary systems, we apply a Doppler shift using the line-of-sight velocity of the barycentre.

\subsection{From single-star spectra to an IFU data cube}

The next step is to define the observational setup of the simulated IFU instrument, in order to obtain the 3-dimensional data cube of our observation. We design an instrument with a field-of-view of $20\times20$ arcsec$^2$ and a spaxel scale of 0.25 arcsec,  similar to the properties of FLAMES@VLT  in ARGUS mode (pixel scale of 0.3 arcsec, field-of-view $\sim7\times4$ arcsec$^2$, resolving power R=10\,000, in the visible range with spectral coverage of 600-1000 \AA; \citealp{Pasquini2002}). Note that these properties can be changed in order to match other instruments, for example, Gemini/GMOS. The covered wavelength range will be $8400-8800$ \AA, with a spectral sampling of 0.1 \AA/pixel. In this particular example, when we place the simulated globular cluster at 10 kpc (Simulation A), a total of $\approx40\,000$ stars fall within the field-of-view of the simulation.

We also need to define a point spread function (PSF) for our observation. We implement a Gaussian PSF, but we will test later the effect of a different PSF shape (Moffat PSF, see 
Sec. \ref{PSFetc}). We set the seeing conditions assigning a FWHM of the PSF of 1 arcsec; typical of ground-based non-adaptive optics assisted observations. After convolving each star 
with the resulting PSF, we sum the Doppler-shifted spectra of all the stars falling in each spaxel, properly weighted by their PSF. In this way for each spaxel we have a spectrum and the 
corresponding luminosity information (obtained by summing up all the flux within each spaxel).

Finally, we add Poisson noise to the final spectra in order to match the desired signal-to-noise ratio $S/N$. The typical $S/N$ values that we consider are $S/N\simeq5$, 10, 20 per {\AA}.\footnote{The $S/N$ reported is the average value per spaxel in the field-of-view; for an average $S/N\simeq10$, the faintest spaxel has a $S/N\simeq3$ and the brightest  $S/N\simeq30$.} In practice, the chosen S/N does not significantly affect the final results, since we will construct the binned radial profiles such that the delivered S/N in each bin is approximately constant (i.e. the size of each bin is chosen to have a fixed S/N).

In the left panel of Fig. \ref{fig:map_lum} we show the luminosity map of the globular cluster at 10 kpc, observed with seeing of 1 arcsec and $S/N\simeq10$ per {\AA}. The luminosity map is constructed summing up all the flux from the spectra within each spaxel, like for real observational data. In the right panel we show the typical spectrum of a single spaxel (both with and without noise).

Note that we do not include any sky lines in addition to the synthetic spectra, since sky subtraction in the Calcium triplet wavelength region is efficiently performed in observational studies \citep{Hanuschik2003}.

\begin{figure*}
\centering
\includegraphics[width=1\textwidth]{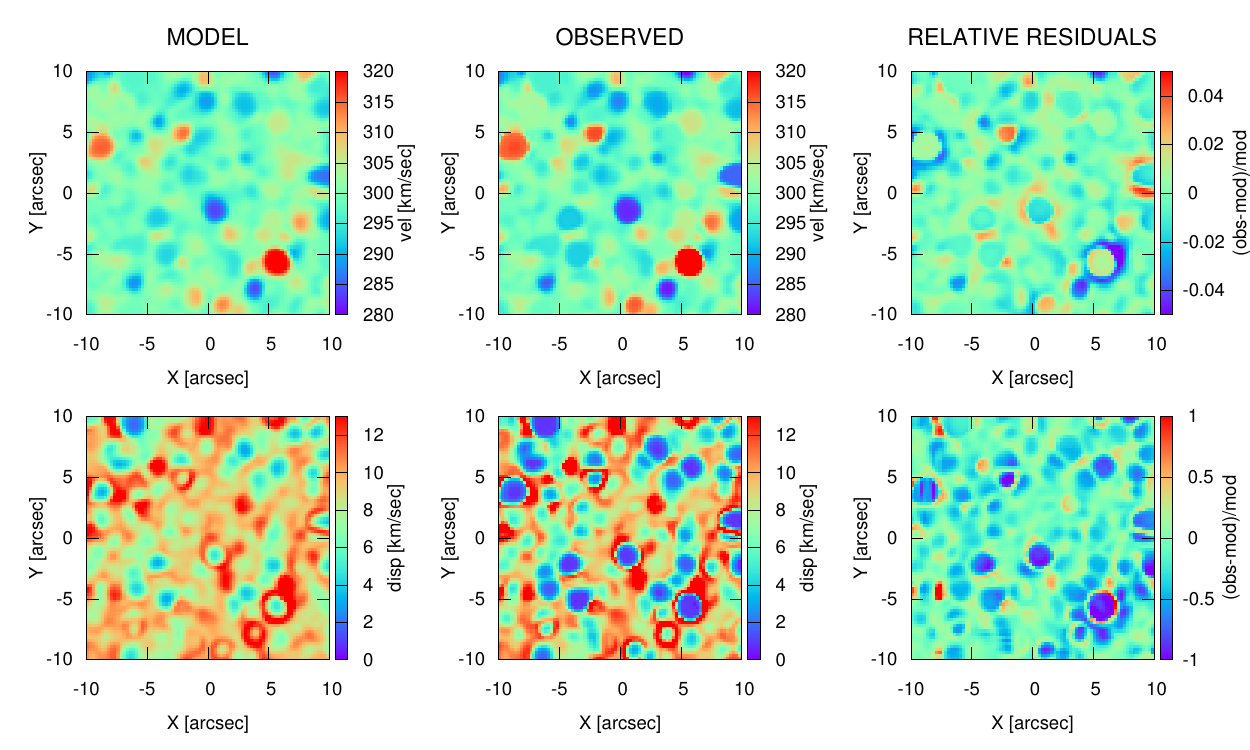}
\caption{\textbf{Top row:} Mean velocity map for one of our IFU mock observations relative to Simulation A with 1 arcsec seeing. The left panel shows the map computed directly from the model, the central panel the map recovered from the mock observation using pPXF to extract the kinematics, and the right panel the relative residuals between observed and model map. \textbf{Bottom row:} Velocity dispersion maps of the model and observed GC, and map of the relative residuals of the two. This mock observation is carried out without considering noise in the measurements, to highlight the intrinsic biases possibly present in our procedure. Our measuring routine allows us to recover the kinematics of the model. The larger discrepancies between model and observations are found in the proximity of the brightest stars for the velocity dispersion measurements (note different plotting ranges for mean velocity and velocity dispersion maps).}
\label{fig:maps_kin}
\end{figure*}
 \subsection{Measuring the kinematics}
 \label{sec:kinematics}
After producing the data cube of our IFU observation, we can proceed with the measurement of the kinematics from the spectra. In common with many observational studies, we use the 
penalised pixel-fitting (pPXF) program of \citet{Cappellari2004} to determine the integrated kinematic properties of each spectrum. This software allows us to obtain the mean velocity and 
the velocity dispersion from the measured spectra, from the shift and from the broadening of the spectral lines.\footnote{Note that pPXF is also capable of measuring higher order moments of the line-of-sight velocity distribution, i.e. $h_3$ and $h_4$. After verifying that the measurements of these higher moments are consistent with zero (as expected for quasi-isotropic stellar systems with no internal rotation), we decided to limit pPXF to fit only for the first two moments, that is the mean velocity and the velocity dispersion.} One of the principal strengths of the pPXF method is the ability to reduce the often significant effects of template mismatch (i.e. the use of template spectra that do not adequately reflect the observed spectra) on the determined kinematics. In order to ensure that template mismatch is negligible we use a total of 16 high-resolution synthetic spectra taken from our simulation (i.e. the original MARCS library spectra before Doppler shifting), making sure they sufficiently cover the parameter space displayed by the stellar types present in our GC simulations. We select representative stars along the colour magnitude diagram: 5 main sequence stars, 8 giant stars and 3 horizontal branch stars. 

With the measured kinematics we are able to construct the mean line-of-sight velocity map and the associated velocity dispersion map. We will refer to these maps as the observed maps. They 
can be directly compared with the model maps. The model map is obtained by calculating for each spaxel the luminosity/PSF-weighted kinematics directly from the Monte Carlo simulation. 
Fig. \ref{fig:maps_kin} shows in the first row the model mean velocity map, the observed mean velocity map and the relative residuals between the two. In the second row, the corresponding velocity dispersion maps are shown. The figure clearly indicates that our measuring routine successfully recovers the average velocity map of the cluster, since our observed maps are consistent with the model ones. 

Fig. \ref{fig:maps_kin} also shows that problems in recovering the internal kinematics of the cluster arise in correspondence to the brightest stars (see for comparison the luminosity map in 
Fig. \ref{fig:map_lum}). In these spaxels, the spectra are completely dominated by the contribution of one or few bright stars, therefore the measured kinematics is biased. 
For example, the measured velocity dispersion in these bright spaxels approaches zero, since by definition the spectrum of one star has no velocity dispersion (other than that caused by
internal line broadening). We will discuss in Sect. \ref{sec:masking} how to minimise this biasing effect introduced by the presence of bright stars.  

Finally, we are able to extract from our data one-dimensional velocity dispersion profiles. We divide our field-of-view in annular bins and sum all the spectra in each bin. With pPXF we measure 
the velocity dispersion from the broadening of the lines of the summed spectra. The advantage of constructing profiles lies in the fact that a higher signal-to-noise per radial bin is reachable, 
making the velocity dispersion measurements more reliable and less affected by stochastic spaxel-to-spaxel noise. The radial size for the annular bins is such that every bin contains the same 
signal-to-noise. We use as the radial position of a bin the radial value at which half of the total number of counts is reached. Note that no error bars are shown in the figures, since the formal errors from pPXF are smaller than the symbols in the graphs. Additionally, we assume the centre of the cluster to be the one given 
by the Monte Carlo cluster simulation and test in Sect. \ref{PSFetc} the effects introduced by a misidentification of the centre.

\begin{figure}
\centering
\includegraphics[width=0.5\textwidth]{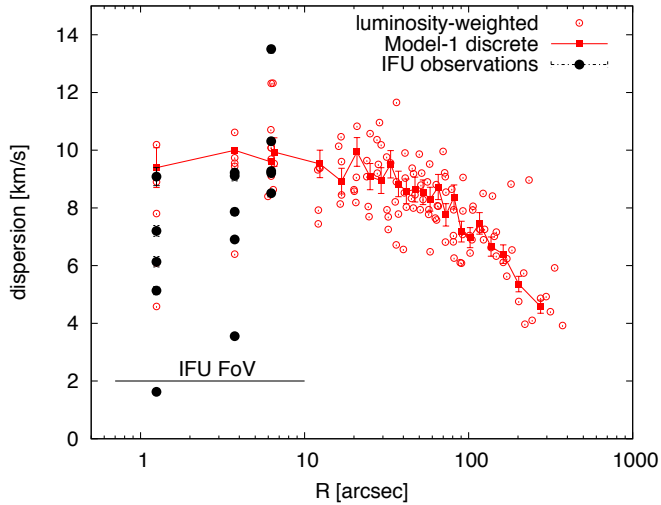}
\caption{Velocity dispersion profiles of the giant stars in our GC model (Simulation A). The open red circles represent the luminosity-weighted velocity dispersion of the five independent realisations of our model
(the bins vary slightly due to stochastic variation); the red squares joined by the red line are the discrete measurements of the velocity dispersion constructed without considering the 
luminosity information of the stars. Each bin contains 200 stars. The black points are instead the measured kinematics from our mock IFU observations of the five realisations, with the horizontal bar indicating the field-of-view (FoV). An intrinsic scatter due to the star-to-star variation of the luminosity is present in the luminosity-weighted profile of our model and this scatter is reflected in the mock observations.}
\label{fig:intrinsic_stoch}
\end{figure}

\section{Luminosity-weighted vs. discrete kinematics}
In this section we will investigate the observational biases present in integrated-light spectroscopic observations. First, we remind the reader that the kinematic information delivered by integrated-light spectroscopy observation is intrinsically luminosity weighted. However, the physical ingredient we are ultimately interested in is not the luminosity, but the mass.

Therefore, using luminosity-weighted kinematics requires some additional caution, since it automatically biases the measured kinematics toward the properties of the brightest stars that 
dominate the spectra, typically red giant branch stars in old stellar systems such as GCs. This can generate two problems: 1) adding a stochastic uncertainty, due to the fact that only few 
bright stars are present in the field-of-view; 2) introducing a systematic bias if stars with different luminosities (and different masses) have different kinematics because 
of mass segregation. We address both of these issues in the following, using the IFU mock observations of the five independent random realisations of our GC (Simulation A), with a fixed seeing condition of 1 arcsec and an average signal-to-noise of $S/N\simeq10$ per {\AA}. From these, we extract the velocity dispersion profiles as described in Sect. \ref{sec:kinematics}.

\begin{figure*}
\centering
\includegraphics[width=1\textwidth]{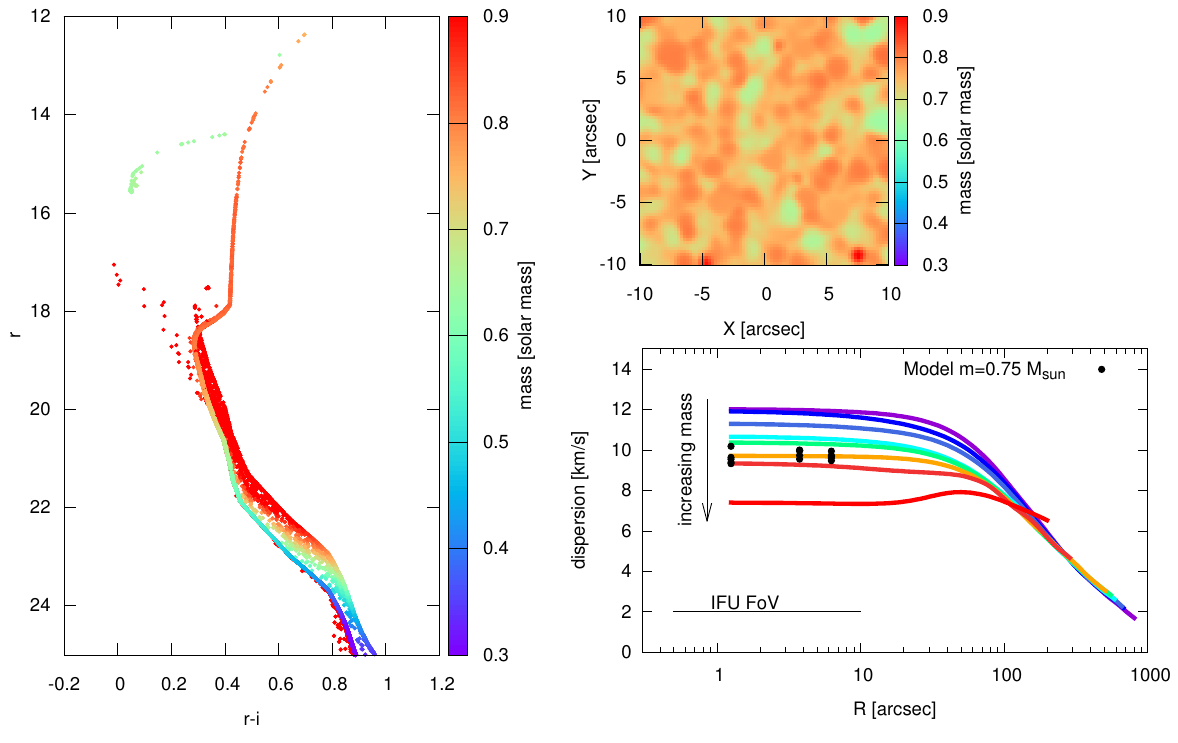}
\caption{\textbf{Left panel:} Colour-magnitude diagram of the stars in the field-of-view (FoV) of our IFU simulation (Simulation A) colour-coded by stellar mass. Objects with masses $>0.9$ M$_{\odot}$ are either stellar binaries or blue stragglers stars. \textbf{Top-right panel:} Luminosity-weighted map of the stellar mass distribution probed by the FoV of our IFU simulation. The typical stellar mass of the kinematic tracers 
that contribute to the spectra in each spaxel is 0.75 $M_\odot$. \textbf{Bottom-right panel:} Velocity dispersion profiles as a function of the projected radius R for stars with increasing mass (from top to bottom). The profiles correspond to 8 stellar mass bins of 0.1 $M_\odot$ width each, with an average stellar mass of $\simeq$ 0.25, 0.35, 0.45, 0.55, 0.65, 0.75, 0.85, 0.95 $M_\odot$, respectively.
The colour scale of the lines is the same as that displayed in the left panel. Stars with lower mass display higher velocity dispersion due to mass segregation and partial energy equipartition. The black dots are the discrete velocity dispersions of the five realisations of our model in the FoV of our IFU 
observation, for stars with mass $0.70-0.80$ $M_\odot$.}
\label{fig:segregation}
\end{figure*}

\subsection{Role of stochasticity}
We illustrate the problem connected to stochasticity in Fig. \ref{fig:intrinsic_stoch}, in which we show the velocity dispersion profile of our model GC constructed using giant stars only (as is typically the case in studies which use measurements of resolved stars). We construct the profile in two different ways: first with discrete data, neglecting the luminosity information (the standard procedure used when discrete line-of-sight velocity measurements are available), then constructing a corresponding luminosity-weighted velocity dispersion profile. While for the former only one model is shown, for the latter we plot the profiles of the five different independent realisations of our GC simulation.\footnote{Little scatter is present in a discrete non-luminosity 
weighted profile.} Both profiles are built with bins containing 200 stars each.
 
The plot clearly shows that, already in the GC model itself, the luminosity-weighted kinematic profiles are influenced by stochastic scatter due to the luminosity differences between stars. Note that, in fact, while all stars in the giant branch have approximately the same mass ($\simeq0.85$ $M_\odot$), their luminosity can vary over 6 magnitudes. This intrinsic stochasticity in the model 
is then transferred to the profiles measured from our mock IFU observations that show a similar amount of scatter. This scatter is high enough to prevent us from obtaining any sound measurement of the velocity dispersion in the central region of the cluster, if we were to use the uncorrected data. We therefore need a way to correctly estimate and minimise this stochasticity present in luminosity-weighted data before attempting any dynamical interpretation of our data. We will quantify the intrinsic scatter and describe the proposed strategy to minimise it in Sect.~\ref{sec:masking}.

\subsection{The true kinematic tracer}
\label{sec:tracers}

Here we investigate which stars are carrying the kinematic information in our IFU observations, that is, for which stars we are able to measure the kinematics. We remind the reader that our GC model incorporates the dynamical effects of mass segregation, that acts to bring the clusters toward (partial) energy equipartition \citep{TrentivanderMarel2013}. This means that the most massive stars in the cluster sink toward the centre while losing energy, while less massive stars gain energy and are preferentially found in the outer regions of the cluster. As a result, higher-mass stars have a lower velocity dispersion with respect to lower-mass stars. This effect is aggravated in the central part of a cluster where mass segregation takes place more efficiently because of the higher density and the shorter relaxation 
time.

It is therefore necessary to know which stars act as the kinematic tracers that we are observing, before interpreting the resulting velocity dispersion measurements.
We therefore construct a luminosity- and PSF-weighted mass map from our mock observations to understand what is the typical value of the stellar mass of the tracers carrying the kinematic information in each spaxel. We present the mass map in the top-right panel of Fig. \ref{fig:segregation} and the colour-magnitude diagram of the stars in the FoV of our simulation (left panel). We show that the average mass of the kinematic tracers is $\simeq0.75$ $M_\odot$ with little scatter ($0.65\lesssim m_\star\lesssim0.85$ $M_\odot$). In the bottom-right panel of Fig. \ref{fig:segregation} we show the interpolated velocity dispersion profiles constructed using the discrete velocities of our GC model for 8 different stellar mass bins (from 0.25 $M_\odot$ to 0.95 $M_\odot$). We also over-plot the points corresponding to the five realisations of our model for the mass bin traced by the IFU observation (0.75 $M_\odot$) in the central region of the cluster corresponding to our IFU field-of-view.

From this analysis we show that the central velocity dispersion strongly depends on which kinematic tracer we are actually measuring, with lower mass stars displaying higher velocity dispersion. In our IFU observations the average stellar mass involved is 0.75 $M_\odot$, slightly lower than the typical giant stars mass of 0.85 $M_\odot$.\footnote{Note that an additional dependence can be introduced by the wavelength range and the specific spectroscopic features that are used to measure the kinematics; for example, spectral features like CO bands will be preferentially sensitive to giant stars \citep{Lanzoni2013}.} The small difference in mass between the red giant stars and the kinematic tracer of our IFU simulation produces a difference in the velocity dispersion profiles that is $<1$ km s$^{-1}$. However, we caution that mass segregation and energy equipartition strongly depend on the evolutionary history of the specific cluster considered.

In the following analysis, when referring to the model velocity dispersion profile we will consider only the profiles built from stars in the mass bin $0.7-0.8$ $M_\odot$, since this is the proper kinematic tracer in our IFU observations.

\begin{figure*}
\centering
\includegraphics[width=1\textwidth]{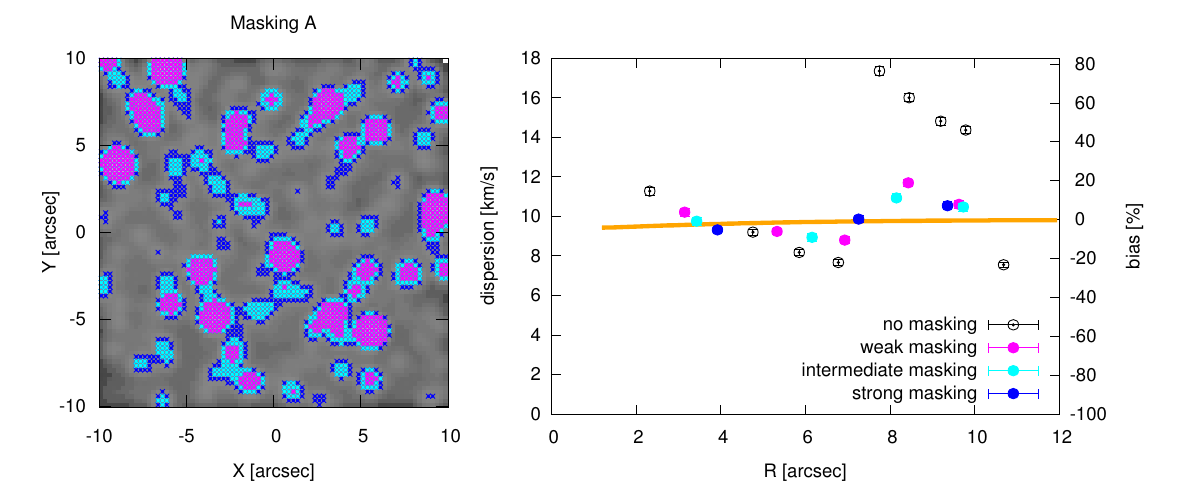}
\caption{\textbf{Left panel:} Luminosity map in logarithmic scale of one of our IFU mock observations (Simulation A, 1 arcsec seeing, $S/N\simeq10$ per {\AA}) with the masked spaxels (masking procedure A) indicated by colours. Magenta spaxels refer to weak masking, cyan to intermediate masking, and blue to strong masking, discarding 10\%, 20\%, and 30\% of the brightest 
spaxels, respectively. \textbf{Right panel:} Velocity dispersion profiles in the field-of-view of the IFU observation for the three maskings of the simulation shown in the left panel. The black open circles are the velocity dispersions computed without masking, while the orange line indicates the expected model velocity dispersion profile (see Figure \ref{fig:segregation}). The vertical axis on the right indicates the bias with respect to the expected velocity dispersion, in percentage. Stronger masking leads to a velocity dispersion profile that progressively approaches the true profile, by reducing the scatter due to bright stars. The profiles are constructed as described in Sect. \ref{sec:kinematics} keeping a constant signal-to-noise per radial bin.}
\label{fig:maskingA}
\end{figure*}

\begin{figure*}
\centering
\includegraphics[width=1\textwidth]{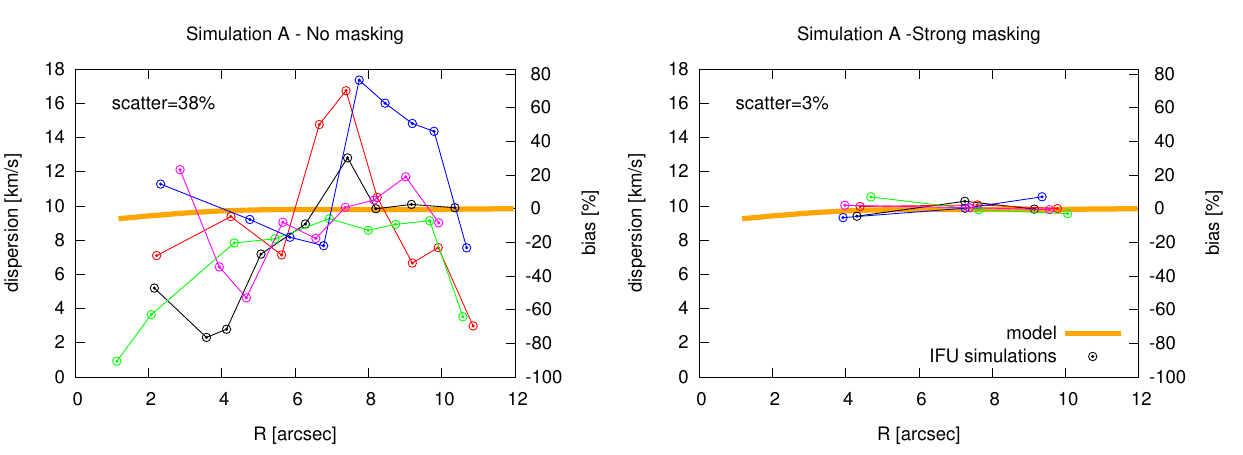}\\
\includegraphics[width=1\textwidth]{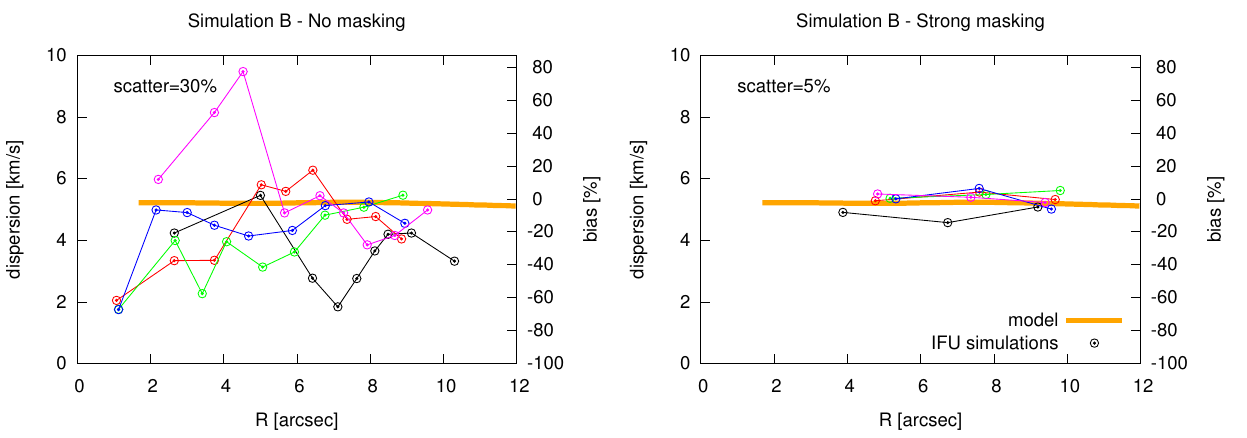}\\
\includegraphics[width=1\textwidth]{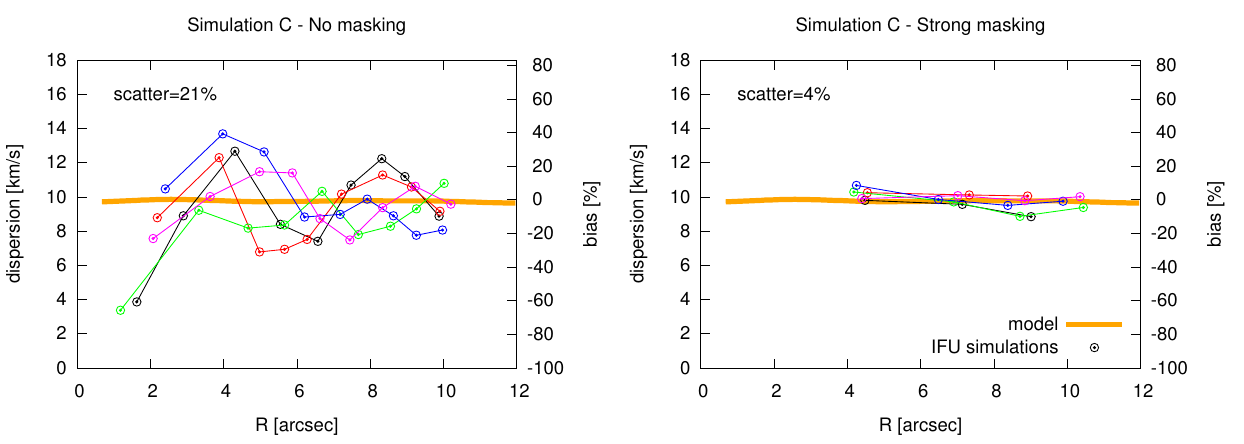}
\caption{Comparison of masked and non-masked velocity dispersion profiles obtained for the 3 sets of IFU simulated observations (Simulation A, B and C; 5 independent realizations each) with 1 arcsec seeing and $S/N\simeq10$ per {\AA}. The profiles have been constructed as described in Sect.\,\ref{sec:kinematics}, keeping a constant signal-to-noise per radial bin. The orange lines indicate the expected model velocity dispersions (see Figure \ref{fig:segregation}) and the vertical axes on the right indicate the bias with respect to the expected velocity dispersion, in percentage. \textbf{Left panels:} Non-masked velocity dispersion profiles showing a strong scatter in velocity dispersion (with respect to the central velocity dispersion) of $\simeq38$, 30, 21\%, for Simulation A, B and C, respectively. Note that in the high-crowding Simulation C, the stochastic scatter of the non-masked profiles is lower than in the case of Simulation A and B, because of the higher number of stars ($\approx4$ times larger) in the FoV.
\textbf{Right panels:} Velocity dispersion profiles corresponding to those in the left panels to which strong masking has been applied (masking A, eliminating 30\% of the brightest spaxels in the field-of-view, see Sect. \ref{sec:masking}). After masking, the profiles accurately reproduce the models and the scatter around the expected values of velocity dispersion is reduced to a few percent.}
\label{fig:masking_non_masking}
\end{figure*}

\section{Analysis of IFU simulations}
\label{sec:masking}

We now describe a procedure to analyze the IFU mock data in order to efficiently recover kinematic information consistent with the actual kinematic tracer involved in the observations, and to 
minimize stochastic scatter in the luminosity-weighted data. In order to do so, it is critical to mask the information conveyed by the bright stars. The masking procedure that we devise, allows us to discard from the analysis those spaxels considered to be significantly contaminated by the light of a single or a few stars. This is a commonly used strategy (e.g., \citealp{Luetzgendorf2011}) and here we intend to adequately understand its efficiency, comparing the results directly with the true model cluster. In this section, we will use all 3 sets of simulations (Simulation A, B and C, see Sect. \ref{sec:MCsim}), in order to test the validity of our analysis for clusters with different concentration and crowding. We use a fixed seeing of 1 arcsec and $S/N\simeq10$ per {\AA}. In Sect. \ref{PSFetc}, we investigate the effect of bad seeing conditions, changing the shape of the PSF and the presumed center of the cluster.

\subsection{Masking of the observations}
\begin{figure*}
\centering
\includegraphics[width=1\textwidth]{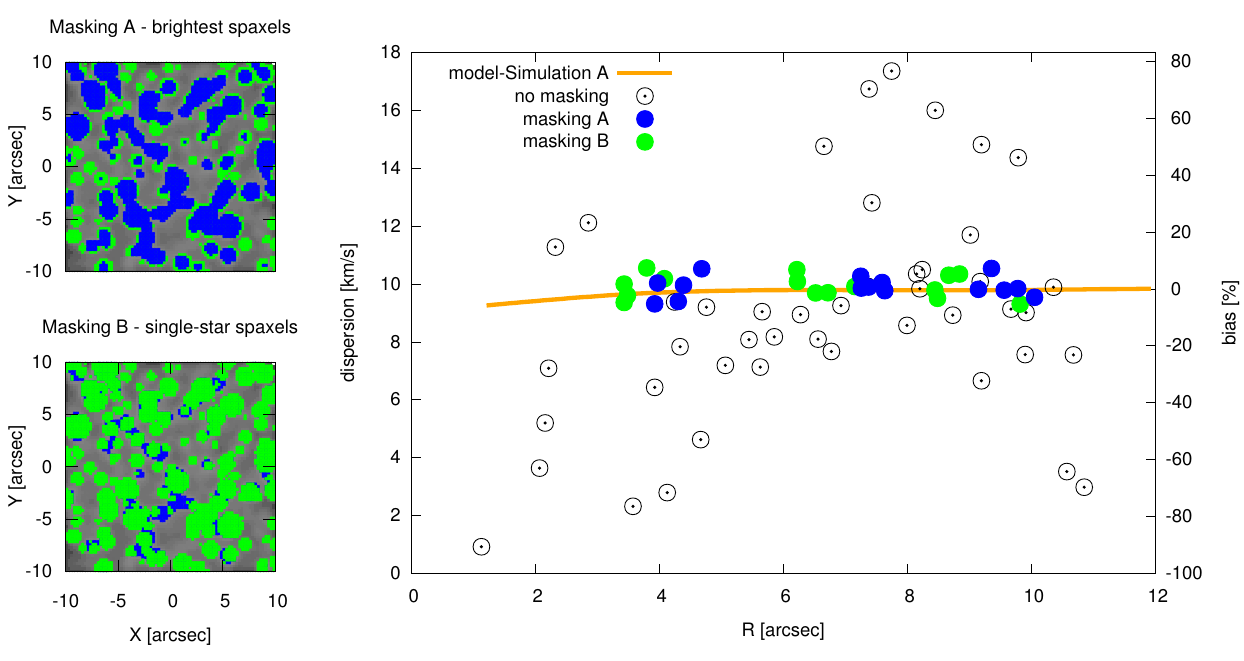}
\caption{\textbf{Left panels:} Luminosity maps in logarithmic scale with masked spaxels indicated in blue for masking strategy A (masking of the 30\% brightest spaxels in the FoV) and in green for masking B (masking of the spaxels dominated by a single bright star), applied to an IFU observation with 1 arcsec seeing,  and $S/N\simeq10$ per {\AA} (Simulation A). \textbf{Right panel:} Velocity dispersion profiles for the five IFU observations without masking (open circles) and after masking according to strategy A (blue points) and B (green points). The orange line indicates the expected model velocity dispersion profile (see Figure \ref{fig:segregation}) and the vertical axis on the right indicates the bias with respect to the expected velocity dispersion, in percentage. The two masking strategies discard approximately the same 
``bad spaxels'' (30-40\% of the total number of spaxels) and yield velocity dispersion profiles fully consistent with each other. Both masking strategies allow for an efficient recovery of the expected model velocity dispersion profile, and significantly reduce the stochastic scatter (from a scatter of $\simeq38$\% around the central velocity dispersion value to $\simeq3\%$).}
\label{fig:masking_comparison}
\end{figure*}

To define a masking strategy, we first need to specify what is a ``bad spaxel'' that is contaminating our data. If we consider that the brightest stars are responsible for biasing our measurements of the integrated kinematics, then the brightest spaxels in our field-of-view can be defined as the ``bad spaxels''. We will refer to this simple masking as Masking A. As an alternative, one could define as ``bad spaxels'' those spaxels whose light comes preferentially from a single bright star. We will refer to this alternative masking as Masking B. Note that the two definitions of ``bad spaxels'' do not strictly coincide, for example a spaxel could be simply very bright because several stars are contributing to its luminosity; it would therefore be considered a ``bad spaxels''  in masking A but not in masking B.

We first consider Masking A and discard from the kinematic analysis the brightest 10\%, 20\%, and 30\% spaxels. We refer to these maskings as weak, intermediate, and strong masking, respectively, and we apply them to each of the five IFU mock observations of our model clusters. The result, for only one of the realizations of Simulation A, is shown in Fig. \ref{fig:maskingA}. The left panel shows the luminosity map superimposed with the ``bad spaxels'' for the three different maskings. The right panel presents the velocity dispersion profiles in the field-of-view of the IFU observation, after the masking has been applied, compared to the profile obtained with no masking, as well as the profile expected directly from the model (see Sect. \ref{sec:tracers} and Fig. \ref{fig:segregation}). The figure shows how a progressively stronger masking ensures recovery of the true model velocity dispersion profile, eliminating the stochastic fluctuations due to bright stars.

Incidentally, we note that, while masking eliminates the brightest spaxels, it does not introduce any bias on the average mass of the kinematic tracers contributing to the remaining spaxels. The average luminosity-weighted mass per spaxel remains constant at $\simeq0.75$ $M_\odot$ even after masking.

In the top row of Fig. \ref{fig:masking_non_masking} we compare the velocity dispersion profiles obtained without masking and the one after applying strong masking to the 5 realizations of the IFU Simulation A. In each panel the velocity dispersion profiles for the five independent realizations are shown. The unmasked profiles are strongly affected both by scatter in the values of velocity dispersion and in the values of radial position. We quantify the scatter in velocity dispersion calculating the biweight standard deviation of all the data points of the velocity dispersion profiles. The initial stochastic scatter is $\simeq38$\% around the central velocity dispersion and is reduced of an order of magnitude when masking is applied, reaching the value of $\simeq3$\%. Applying the masking we are both able to accurately recover the values of velocity dispersion expected from the model (see Sect. \ref{sec:tracers}) and to considerably reduce the scatter.

We apply the same analysis also to Simulation B and Simulation C, the former providing a higher concentration cluster and the latter a FoV with higher crowding (see Sect. \ref{sec:MCsim}). As shown in the middle row (Simulation B) and bottom row (Simulation C) of Fig. \ref{fig:masking_non_masking}, in both cases the masking procedure allows us to recover the values of velocity dispersion expected from the models and significantly reduce the scatter. Note, in particular, that for the high-crowding Simulation C, the initial stochastic scatter is $\simeq21\%$, less severe than in the corresponding case of Simulation A and B. This can be explained by the higher number of stars in the FoV of our IFU mock observations, connected to the higher central luminosity density of the simulated cluster ($\approx4$ times more stars than in Simulation A, consistent with the fact that Simulation C is Simulation A observed at twice the distance, 20 kpc). This reduces the shot noise due to low number statistics in agreement with what reported in \citet{Dubath1997} for integrated-light slit spectroscopy. After masking is applied, the scatter reaches a value of $\simeq4$\%, comparable to the value reported for the case of the less dense Simulation A.

In the same way as described above, we test the masking strategy B. We discard those spaxels for which at least one single star contributes more than 50\%, 40\%, 30\% of the luminosity. We refer to these maskings as weak, intermediate, and strong masking, respectively. In Fig. \ref{fig:masking_comparison} we compare the results of the two different masking strategies applied to Simulation A. The panels to the left show the luminosity map with the masked spaxels indicated in blue and in green, for Masking A and Masking B, respectively. In this figure we consider only the strongest masking flavors: in the case of Masking A, exactly 30\% of spaxels are discarded, while for Masking B approximately 40\% of spaxels are removed. Note however that, in this case, the two masking strategies agree with each other, since they both remove nearly the same ``bad spaxels''. Moreover, the right panel of the figure shows that the resulting velocity dispersion profiles for the five IFU observations obtained using the two masking strategies are fully consistent with each other.

A quantitatively similar result is obtained when comparing the two masking techniques applied to the higher concentration Simulation B. For the high-crowding Simulation C, although the two masking approaches lead to quantitatively comparable velocity dispersion profiles, Masking B proves more efficient (i.e., it removes less spaxels, $\approx20\%$ instead of 30\% in Masking A). This can be explained by the fact that Masking A eliminates also those spaxels that are bright due to the contribution of many stars, that are therefore misidentified as ``bad spaxels''.

In the following section we summarise the main limitations of the two masking techniques discussed here.

\begin{figure*}
\centering
\includegraphics[width=1\textwidth]{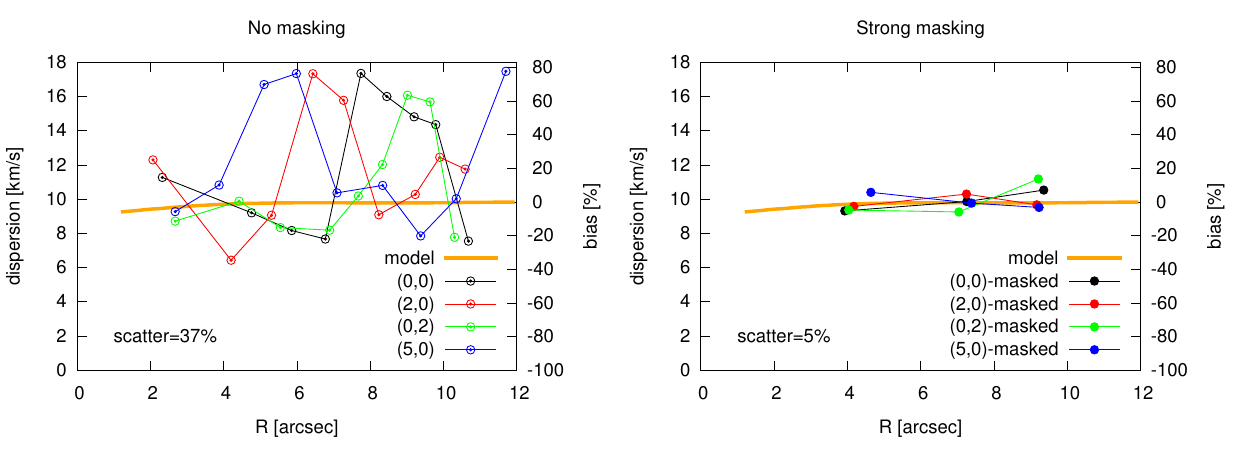}
\caption{Comparison of the velocity dispersion profiles obtained from one IFU simulated observation using 4 different centers for the cluster (Simulation A); in particular no offset from the default center given by the simulation (black line), 2 arcsec offset along the x-axis (red line), 2 arcsec offset along the y-axis (green line), 5 arcsec offset along the x-axis (blue line). The orange line indicates the expected model velocity dispersion (see Figure \ref{fig:segregation}) and the vertical axis on the right indicates the bias with respect the expected velocity dispersion, in percentage. \textbf{Left panel:} no masking applied. \textbf{Right panel:} masking of ``bad spaxels'' has been applied (strong masking A, see Sect. \ref{sec:masking}). The shape of the velocity dispersion profiles strongly depends on the choice of the center; however the correct value of velocity dispersion is recovered after applying masking, since the scatter is significantly reduced from $\simeq37$\% to $\simeq5$\%.}
\label{fig:center}
\end{figure*}

\subsection{Limitations of the masking approach}
We showed in the previous section that, in the cases tested here, masking allows the reliable recovery of the expected unresolved kinematics. However, this procedure rejects from the analysis a fairly large amount of bright and high signal-to-noise spaxels. It is therefore important to highlight the main limitations of masking and in particular those connected to the two masking strategies used in this work.
\begin{itemize}
\item Masking enforces a loss of spatial resolution, since a relatively high number of ``bad spaxels'' are removed. Therefore some information on the central kinematic properties of a cluster is unavoidably not recoverable (note that in Fig. \ref{fig:maskingA} - \ref{fig:masking_comparison} the masked profiles do not sample the inner $\lesssim4$ arcsec, corresponding to $\approx$3-4 PSF elements around the center). 
\item Masking A removes only the brightest spaxels in the FoV, leaving in the analysis spaxels that are still dominated by the light of a single star, but that are not bright. This drawback is not particularly severe if velocity dispersion profiles are constructed (i.e. the FoV is divided in annular bins and a summed spectrum is calculated for each bin). In fact, since these spaxels are faint, their relative contribution to the summed spectra is low. However, discarding these spaxels is crucial if a spaxel-by-spaxel kinematic map needs to be constructed. Moreover, in a very crowded FoV, Masking A would misidentify as ``bad spaxels'' also those spaxels that are very bright due to the contribution of the light of many less bright stars.
\item Masking B properly identifies the spaxels dominated by a single star only. However, it requires additional observational information (e.g. distribution of the stars in the FoV and the modeling of the PSFs), to determine the relative contribution of the stars in each spaxel.
\item We showed that in the case of a less crowded FoV (Simulation A), Masking A performs more efficiently than Masking B (i.e. it delivers the expected velocity dispersion profiles, while removing a lower number of spaxels). For a more crowded FoV (Simulation C) Masking B is more efficient.
\end{itemize}

Moreover, we wish to stress that the high signal-to-noise spaxels dominated by a single star that are rejected with the masking approach still contain important kinematic information that can be used in complementary ways, for example, to determine discrete (i.e., individually resolved) kinematics (see for example \citealp{Kamann2013,Kamann2014,Lanzoni2013}). Finally, we tested an alternative approach to masking, consisting in performing the kinematic analysis after normalising the luminosity of each spaxels, so that every spaxel has the same relative contribution in the final kinematic measurements (as performed in \citealp{Noyola2010}). Our results show that this procedure leads to an effect slightly worse than what obtained with a weak masking (see Fig. \ref{fig:maskingA}).

\subsection{Shape of the PSF,  bad seeing conditions, and misidentification of the center of the cluster}
\label{PSFetc}

In this section we discuss the general validity of our results, exploring different parameter configurations for our IFU mock observations. In particular, we explore the changes introduced in case of a different shapes of the PSF, bad seeing conditions and slight mis-identifications of the cluster centre.

First, we test the effect of modeling the shape of the PSF using a \citet{Moffat1969} light distribution instead of a Gaussian PSF. We fix a seeing of 1 arcsec (FWHM) and model the Moffat distribution with a shape-parameter $\beta=2.5$ \citep{Trujillo2001}. Although the PSF appears now more extended (due to the wings of the distribution), the masking techniques described above still recovers the expected values of velocity dispersion.

We then test a case of bad seeing conditions of 2 arcsec, using a Moffat PSF for the high-crowding simulation (Simulation C). Before applying masking, the five independent realisations show a stochastic scatter of 12\%. This value is significantly lower than the corresponding case with 1 arcsec seeing (scatter of 21\%),\footnote{This does not mean that bad seeing conditions are best suited for integrated-light spectroscopy studies. In fact, the typical field of view of many available IFU spectrographs is of a few arcseconds only (i.e., comparable to or slightly larger than the seeing considered here) and if just a couple of bright stars are present, their light would completely dominate the acquired spectra.}
since now the light of the stars is distributed over several spaxels because of the more spatially extended PSF shape. When strong masking is applied, the scatter is only slightly reduced to 8\%, hinting that masking is not particularly efficient in cases of bad seeing.   

As a final test we explore the effect of changing the identification of the centre of our simulation (using Simulation A). We shift the centre by 2 or 5 arcsec along the x-axis, or 2 arcsec along the y-axis, and we construct the velocity dispersion profile with and without applying the masking procedure outlined. The adopted shifts correspond to $7\%-20\%$ the value of the core radius of $R_c\simeq1.3$ pc. The result is shown in Fig. \ref{fig:center}, where it is clear that even a small change of the centre causes a strong variation of the velocity dispersion profile: we estimate a scatter of $\simeq37$\% around the expected value of velocity dispersion. This is due to the changed position of the few bright stars with respect to the binning used. However, after masking (right panel Fig. \ref{fig:center}) we are able to recover the expected velocity dispersion profile, since we remove the bright stars that were biasing our result, reducing significantly the scatter to $\simeq5$\%.
Note that the issues connected to the misidentification of the centre may change when considering a simulation with a central IMBH.

\section{Conclusions}

We have constructed realistic integral field spectroscopic mock observations of the central region of globular clusters. Our software SISCO (Simulating IFU Star Clusters Observations) 
produces a data cube with spectra and luminosity information in a given wavelength region (e.g., the Calcium triplet region, $8400-8800$ \AA) for every spaxel in a desired field-of-view, with adjustable seeing conditions and signal-to-noise. The starting point of our mock observations can be any realistic single stellar population cluster model, for which the stellar parameters are given as an output.

Here we applied SISCO to Monte Carlo cluster simulations in which the stellar initial mass function, stellar evolution, initial binary fraction, a realistic number of stars (up to $N\approx2\,000\,000$) and realistic concentrations are taken into consideration. No IMBHs are considered in this work. We used the output of our mock observations to extract the internal kinematics of the cluster from the Doppler shift and broadening of the spectra. From these we extract kinematic maps and velocity dispersion profiles in the same manner as observational studies.

With the mock observations we aim to understand the biases resulting from using integrated-light spectroscopy to measure the kinematics of partially resolved stellar systems. This is a first step to understand the discrepancies reported in the literature between resolved discrete kinematics and unresolved luminosity-weighted kinematics, connected to the detection of IMBHs. From the analysis of our specific set of simulations we find that:

\begin{itemize}
\item The luminosity-weighted kinematics from IFU observations can be strongly biased by the presence of a few bright stars. The kinematic data are strongly affected by stochasticity, 
and this prevents reliable measurements of the central velocity dispersion of GCs. Using five independent realizations of a given cluster simulation we estimate that the intrinsic scatter around the expected value of velocity dispersion can be as high as 40\% for the less crowded simulations (central luminosity density of $\simeq60$ L$_{\odot}$ arcsec$^{-2}$) and $20\%$ for the more crowded ones (central luminosity density of $\simeq200$ L$_{\odot}$ arcsec$^{-2}$), in typical seeing conditions. An additional source of stochasticity (quantitatively comparable to the former) is introduced by the particular choice of the center of the cluster, since changing the center position has the effect of changing the positions of the bright stars with respect to the radial bins.

\item The internal kinematics in the central region of a GC depends on the kinematic tracers involved in the observations. High-mass stars have a lower velocity dispersion than low-mass stars 
because of mass segregation. We show that the average kinematic tracers of our IFU (and likely most actual) observations have a mass of $\simeq0.75$ $M_\odot$, slightly lower that the 
typical mass of resolved line-of-sight velocity measurements (giant stars with mass $\simeq0.85$ $M_\odot$). Understanding which tracers are carrying the kinematic information is a necessary 
step when combining kinematic data sets obtained with different and complementary observational strategies. This conclusion will become particular important when using proper 
motions and line-of-sight velocities in the same kinematic analysis. For example, Hubble Space Telescope proper motion measurements sample the kinematics for stars with different masses 
along the main sequence \citep{Bellini2014,Watkins2015}, while line-of-sight velocities sample bright (and more massive) stars only.

\item Given these findings, we are able to assess that luminosity-weighted kinematics is highly dependent on the presence of a few bright stars that can bias (both overestimating and underestimating) the measurements of the central velocity dispersion. This is a first step to explain the discrepancies reported in the literature between resolved discrete kinematics and unresolved luminosity-weighted kinematics. For a sensible comparison of these different measurements, it is necessary to apply a proper treatment to the unresolved integrated-light kinematic measurements, like masking techniques. Note, however, that simulations specifically designed to match particular cases are needed to further understand in details the discrepancies reported in the literature (e.g. see the case of NGC 6388, \citealp{Luetzgendorf2011,Luetzgendorf2015,Lanzoni2013}).

\item We show that, for the specific simulations used in this paper, masking of spaxels contaminated by bright stars allows us to recover measurements consistent with the model of the kinematic tracers.
Moreover, masking reduces significantly the intrinsic scatter around the expected value of velocity dispersion in the luminosity-weighted data, bringing it down to the value of a few percent. Reducing the scatter to a low level is essential to allow for a reliable interpretation of the presence/absence of IMBHs. We report that the efficiency of masking depends on the crowding of the field-of-view and the seeing conditions. In particular, for a highly crowded field-of-view (luminosity density of $\simeq200$ L$_{\odot}$ arcsec$^{-2}$) and bad seeing conditions masking proves to be less efficient. The main limitation of masking is to cause a loss of spatial resolution, typically across the central 3-4 PSF elements. Therefore masking can be used as a complementary approach to the one of extracting the kinematics of only the bright, high signal-to-noise spaxels dominated by single stars.
\end{itemize}

In a subsequent paper we will apply our program SISCO to state-of-the-art dynamical simulations of globular clusters models in which additional physical ingredients are included. In particular, 
we will compare mock observations of models with and without IMBHs to help understand if the presence of an IMBH can be inferred with the application of standard dynamical 
modeling approaches.

\section*{Acknowledgments}
We are grateful to Jonathan Downing for providing the Monte Carlo cluster simulations used in this work. We thank Barbara Lanzoni and Nora L\"utzgendorf for useful discussions and the referee for constructive remarks that have helped improve the quality of the paper. PB acknowledges financial support from Heidelberg Graduate School for Fundamental Physics.

\bibliographystyle{mn2e} 
\bibliography{biblio} 


\end{document}